\documentclass[prd,superscriptaddress,twocolumn,aps,amsmath,amssymb,showpacs,nofootinbib]{revtex4-2}

\usepackage{amsmath}
\usepackage{graphicx,enumerate,xcolor,enumitem,xspace,braket,mathrsfs} % Required for inserting images
\usepackage[colorlinks,citecolor=blue]{hyperref}
\usepackage{comment}
\usepackage{multirow}

\newcommand{\nn}{\nonumber\\}
\newcommand{\tr}{\operatorname{Tr}}

\newcommand{\calM}{\mathcal{M}}

\newcommand{\customsection}[1]{\section{\bf #1}}

\usepackage[normalem]{ulem}

\begin{document}
{\hfill PITT-PACC-2408}
\title{Quantum Tomography at Colliders: With or Without Decays}

\author{Kun Cheng}
\email{kun.cheng@pitt.edu}
\affiliation{PITT PACC, Department of Physics and Astronomy,\\ University of Pittsburgh, 3941 O’Hara St., Pittsburgh, PA 15260, USA}

\author{Tao Han}
\email{than@pitt.edu}
\affiliation{PITT PACC, Department of Physics and Astronomy,\\ University of Pittsburgh, 3941 O’Hara St., Pittsburgh, PA 15260, USA}

\author{Matthew Low}
\email{mal431@pitt.edu}
\affiliation{PITT PACC, Department of Physics and Astronomy,\\ University of Pittsburgh, 3941 O’Hara St., Pittsburgh, PA 15260, USA}

\date{\today}

\begin{abstract}
  In constructing quantum states at colliders, most efforts have focused on utilizing the particle decays to infer the spins to reconstruct the quantum density matrix. We show that the production kinematics of the particles provides sufficient information about the spins to establish quantum tomography event-by-event without using the decays.  We lay out the advantages  of using this ``kinematic approach'' relative to the usual ``decay approach.''   Since the kinematic approach leverages the simplicity of scattering kinematics, this approach promises to achieve optimal statistical results for quantum tomography at colliders.
\end{abstract}

\maketitle

%%%%%%%%%%%%%%%%%%%%%%%
\customsection{Introduction}

The ways that we understand the microscopic world are based on quantum mechanics.  Two notable phenomena that quantum systems can exhibit -- which classical systems cannot -- are entanglement and Bell inequality violation.  Recently these phenomena have been studied at high energy colliders where the quantum states are the spin states of the particle system \cite{Afik:2020onf,Fabbrichesi:2021npl,Barr:2021zcp}.  For instance, in $t\bar{t}$ production at the Large Hadron Collider (LHC), the $t$ and $\bar{t}$ are treated as qubits leading to a bipartite qubit system~\cite{Severi:2021cnj,Afik:2022kwm,Aguilar-Saavedra:2022uye,Fabbrichesi:2022ovb,Dong:2023xiw,Han:2023fci,Barr:2024djo}. Depicting the full properties of this quantum state is known as quantum tomography.

Traditionally, quantum tomography is a procedure that involves performing a complementary set of quantum measurements on an ensemble of identically-prepared quantum states.  Different measurements collapse the states differently and provide complementary information about the quantum state.  At colliders, typically the spin state of a particle is used as the qubit. The spins of particles produced at high-energy colliders are not directly measured but are inferred from the directions of decay products.  A second distinctive feature of quantum information at colliders is that ``fictitious states''~\cite{Afik:2022kwm,Cheng:2023qmz,Cheng:2024btk} are reconstructed when using event-dependent basis, rather than genuine quantum states when using a fixed basis.

In this Letter, we advocate a new method in collider quantum tomography: the ``kinematic approach'', in contrast to the well-known ``decay approach.'' An important advantage of the kinematic approach is that it does not rely on particle decays which enables the use of particles that do not decay in collider experiments.  We perform a comparative study for the advantages and disadvantages of using these two approaches and highlight the complementarity between them.  Since the kinematic approach leverages the simplicity of scattering kinematics, this approach may achieve the optimal statistical results for quantum tomography at colliders.

%%%%%%%%%%%%%%%%%%%%%%%
\customsection{Two Approaches}
\label{sec:formalism}

The fundamental quantity that captures the relevant information about a quantum state is the density matrix $\rho$.  In this work, for the sake of simplicity, we use the two-qubit system, which has a $4 \times 4$ density matrix, but the formalism applies equally to any quantum system.

For the two-qubit system, the density matrix has 15 real parameters and it is most conveniently parametrized as
\begin{equation} 
\label{eq:rho}
\begin{aligned}
\rho = \frac{1}{4}
\bigg(  \mathbb{I}_4
& + \sum_i B^+_i \sigma_i \otimes \mathbb{I}_2
+ \sum_j B^-_j \mathbb{I}_2 \otimes \sigma_j  \\
& +  \sum_{ij} C_{ij} \sigma_i \otimes \sigma_j
\bigg)  ,
\end{aligned}
\end{equation}
where $B^+_i$ is the net polarization of the first qubit, $B^-_j$ is the net polarization of the second qubit, and $C_{ij}$ is the spin correlation matrix.

At high energy colliders, we assume our system is described by a quantum state and then ask questions about this quantum state, such as the extent of their entanglement and whether it is Bell local or non-local.  In effect, by measuring the quantum state we are really performing a fit to a parametrization of the density matrix~\cite{Han:2023fci}.

With this perspective in mind, we show that there are two natural sets of parameters to describe the density matrix. We label the use of the first set of parameters as the {\bf decay approach}.  This is the standard technique in past works \cite{Afik:2020onf,Fabbrichesi:2021npl,Barr:2021zcp}.  In this Letter, we introduce the other set of parameters and label their usage as the {\bf kinematic approach}, which is a distinct method.  This only depends on the $2\to 2$ production kinematics and offers unique advantages to quantum tomography at colliders.  The essential quantum information, contained in $\rho$, may be probed either via the kinematics of production or via the angular distributions of decay products.

%%%%%%%%%%%%%%%%%%%%%%%
\subsection{The Decay Approach} 

In the decay approach, the parametrization of Eq.~\eqref{eq:rho} is taken as the fundamental description of the quantum state.  Quantum tomography is therefore a 15-parameter fit of the form 
\begin{equation} \label{eq:rho_decay}
\rho = \rho(B^+_1, B^+_2, B^+_3, B^-_1, B^-_2, B^-_3, C_{11}, \ldots).
\end{equation}
The parameters of Eq.~\eqref{eq:rho_decay} cannot be measured in a single event but can be extracted from the distribution of angles of decay products.

When a qubit decays, each outgoing daughter particle $a$ has an associated spin analyzing power $\kappa^a$ that quantifies how correlated the angle of that particle's momentum is with the spin of the mother qubit.  The relevant angle $\theta_i^a$ is the angle between the daughter's three-momentum and the axis $i$ in the rest frame of the qubit.  The distributions of $\theta_i^a$ are universal, predicted by a given quantum theory, and allow a statistical measurement of each parameter
\begin{align}
B^+_i &= \frac{3}{\kappa^a}  \langle \cos\theta_i^{a} \rangle, \quad 
B^-_j = -\frac{3}{\kappa^b}  \langle \cos\theta_j^{b} \rangle, \\
C_{ij} &= -\frac{9}{\kappa^{a}\kappa^{b}}
\langle \cos\theta^{a}_i \cos\theta^{b}_j \rangle   ,\label{eq:Cij_decay}
\end{align}
where $a$ and $b$ label the daughter particles of each qubit.  For instance, the charged lepton $\ell$ in the decay of a top quark has the largest spin analyzing power of $\kappa^\ell=1$. 

The decay approach fits the spin correlation matrix without any knowledge of the production.  In principle, other than specifying the system as bipartite, no process-specific information is required to measure $\rho$ and subsequently compute any quantum quantity.

In practice, we often use process-specific theory knowledge to simplify the measurement and calculation of entanglement and Bell inequalities.  For example, in the $t\bar{t}$ system we reasonably assume that the production is unpolarized, CP-invariant, and symmetric around the beamline.  This sets both polarization vectors to zero and reduces the 9 parameters of Eq.~\eqref{eq:rho_decay} to the 4 parameters: $C_{11}$, $C_{22}$, $C_{33}$, and $C_{13}$.\footnote{Throughout this Letter we work in the helicity basis where the indices are $\{1,2,3\} = \{k,n,r\}$.}  With theoretical input, the density matrix is reduced to $\rho=\rho(C_{11}, C_{22}, C_{33}, C_{13})$.  For this simplified system, the concurrence $\mathcal{C}$ and the Bell variable $\mathcal{B}$ are given as\footnote{This closed form of concurrence and Bell variable only applies to $t\bar{t}$ in the high-$p_T$ region, which is entangled as a spin triplet.  For other scenarios there are other closed forms~\cite{Afik:2020onf}.  For systems beyond two qubits it is common to compute a bound on the concurrence rather than the actual value.} 
\begin{align} \label{eq:conc_decay}
\mathcal{C} &= \frac{1}{2}(C_{11} + C_{33} - C_{22} - 1), \\ 
 \label{eq:bell_decay}
\mathcal{B} &= \sqrt{2} (C_{33} - C_{22}).
\end{align}
The conditions for quantum entanglement and Bell inequality violation are
\begin{equation}
\mathcal{C} > 0,
\quad\quad\quad \mathcal{B} > 2,
\end{equation}
respectively. 

From Eq.~\eqref{eq:Cij_decay}, the statistical uncertainty $\Delta C_{ij}^{{ \rm stat}}$ on the measurement of an entry of the spin correlation matrix $C_{ij}$, which follows the normal  distribution for a large number of events $N$, takes the universal form of
\begin{equation}
\Delta C_{ij}^{{ \rm stat}} = \frac{3}{\sqrt{N}} .
\end{equation}
%
%\sout{where $N$ is the number of observed events.}
%
When the concurrence and Bell variable are linear functions of elements of the spin correlation matrix, as in Eqs.~\eqref{eq:conc_decay} and~\eqref{eq:bell_decay}, the uncertainties on $\mathcal{C}$ and $\mathcal{B}$, respectively, are
\begin{align}
\Delta \overline{\mathcal{C}}^{ \rm stat} &= \frac{1}{2} \sqrt{ (\Delta C_{11}^{{ \rm stat}})^2
+ (\Delta C_{22}^{{ \rm stat}})^2 + (\Delta C_{33}^{{ \rm stat}})^2}, \\
\Delta \overline{\mathcal{B}}^{{ \rm stat}} &= \sqrt{2} \sqrt{ (\Delta C_{22}^{{ \rm stat}})^2 + (\Delta C_{33}^{{ \rm stat}})^2}.
\end{align}
where the statistical uncertainties of different $C_{ij}$'s are approximately independent~\cite{SeveriThesis}. 
Anticipating improvements in future measurements, we can define the percentage accuracy for these measurements as
\begin{align}
\label{eq:prec_c_decay}
\frac{\Delta \overline{\mathcal{C}}^{{ \rm stat}}}{\overline{\mathcal{C}}} &= \frac{3\sqrt{3}}{(C_{11}+C_{33}-C_{22}-1)} \frac{1}{\sqrt{N}}, \\
\label{eq:prec_b_decay}
\frac{\Delta \overline{\mathcal{B}}^{{ \rm stat}}}{\overline{\mathcal{B}}} &= \frac{3\sqrt{2}}{C_{33} - C_{22}} \frac{1}{\sqrt{N}}.
\end{align}

%%%%%%%%%%%%%%%%%%%%%%%
\subsection{The Kinematic Approach}

The density matrix can alternatively be parameterized by the kinematic quantities of the scattering angle $\Theta$ of the outgoing particle with respect to the incoming particle and the final particle speed $\beta$ in the center-of-mass frame of the two-qubit system. In this case the density matrix is measured via a 2-parameter fit of the form
\begin{equation}\label{eq:rho_kin}
\rho = \rho(\Theta,\beta).
\end{equation}
The form of Eq.~\eqref{eq:rho_kin} is process-dependent, but calculable once the process is specified.  For example, consider $q\bar q \to t\bar t$.  The polarization of each qubit is zero and the spin correlation matrix takes the form
\begin{equation}\label{eq:Cijqq}
C_{ij} = 
\begin{pmatrix}
\frac{2c_\Theta^2 +\beta^2s_\Theta^2}{2-\beta^2s_\Theta^2} &  0 & -\frac{2c_\Theta s_\Theta\sqrt{1-\beta^2}}{2-\beta^2 s_\Theta^2}\\
        0&\frac{-\beta^2 s_\Theta^2}{2-\beta^2 s_\Theta^2}&0\\
        -\frac{2c_\Theta s_\Theta \sqrt{1-\beta^2}}{2-\beta^2s_\Theta^2} & 0 & \frac{(2-\beta^2)s_\Theta^2}{2-\beta^2 s_\Theta^2}  
    \end{pmatrix},
\end{equation}
where $c_\Theta = \cos\Theta$ and $s_\Theta = \sin\Theta$.  The concurrence $\mathcal{C}$ can thus be measured event-by-event from the two kinematic variables $\Theta$ and $\beta$:
\begin{equation}
\begin{aligned}
\mathcal{C}(\Theta,\beta)
& = \frac{C_{11}(\Theta,\beta) + C_{33}(\Theta,\beta) - C_{22}(\Theta,\beta)-1}{2},  \\
& = \frac{\beta^2 s_\Theta^2}{2-\beta^2 s_\Theta^2}. 
\end{aligned}
\end{equation}
The Bell variable is
\begin{equation}
\begin{aligned}
\mathcal{B}(\Theta,\beta)& = \sqrt{2} | C_{33}(\Theta,\beta)  - C_{22}(\Theta,\beta)| \\ 
& = \frac{2\sqrt{2}s_\Theta^2}{2-\beta^2 s_\Theta^2}.
\end{aligned}
\end{equation}
The measured values of concurrence and the Bell variable are, respectively, the mean of the variables $\mathcal{C}$ and $\mathcal{B}$ 
\begin{equation}
\label{eq:conc_kin}
\overline{\mathcal{C}}= \langle \mathcal{C}(\Theta,\beta) \rangle, \quad 
\overline{\mathcal{B}} = \langle \mathcal{B}(\Theta,\beta) \rangle.
\end{equation} 
With sufficient large number of events, the mean values of $\mathcal{C}(\Theta,\beta)$ and $\mathcal{B}(\Theta,\beta)$ follow normal distributions and the statistical uncertainties of concurrence $\Delta \mathcal{C}$ and the Bell variable $\Delta \mathcal{B}$ are
\begin{equation}
\Delta \overline{\mathcal{C}}^{{ \rm stat}} = \frac{\sqrt{\text{Var}(\mathcal{C}(\Theta,\beta))}}{\sqrt{N}}, \quad
\Delta \overline{\mathcal{B}}^{{ \rm stat}} = \frac{\sqrt{\text{Var}(\mathcal{B}(\Theta,\beta))}}{\sqrt{N}}, 
\label{eq:dvar}
\end{equation} 
where ${\rm Var}(\mathcal{R}) = \langle ( \overline{\mathcal{R}} - \mathcal{R} )^2 \rangle$, $\mathcal{R}$ is the ``random variable'' and $\overline{\mathcal{R}}$ is its mean value.
% where ${\rm Var}(F)$ is the variance of distribution $F$.\footnote{{Since each event uses $\Theta$ and $\beta$ to reconstruct $\mathcal{C}$, the random variable for the variance of $F_\mathcal{C}$ is $\mathcal{C}$.} \Tao{(then why not just replace $F_C \to C$?)}\ML{That's what we originally had. 
 % We can add a note to say that $\mathcal{C}(\theta,\beta)$ is the distribution and $\mathcal{C}$ is the mean value.}} 
The explicit forms of $\mathcal{C}$ and $\mathcal{B}$ regarding kinematic variables are process-dependent.  

By Popoviciu's inequality, the percentage accuracies are bounded by
\begin{equation}
\frac{\Delta \overline{\mathcal{C}}^{{ \rm stat}}}{\overline{\mathcal{C}}}  
\leq \frac{1}{2 \overline{\mathcal{C}}} \frac{1}{\sqrt{N}}, \ \  
\frac{\Delta \overline{\mathcal{B}}^{{ \rm stat}}}{\overline{\mathcal{B}}} 
\leq \frac{\sqrt{2}}{\overline{\mathcal{B}}} \frac{1}{\sqrt{N}}.
\hfill
\label{eq:rel}
\end{equation} 
Compared with Eqs.~\eqref{eq:prec_c_decay} and~\eqref{eq:prec_b_decay}, the statistical uncertainties in the kinematic approach are substantially smaller than in the decay approach. This is because the kinematic approach exploits the full kinematic dependence in measuring quantities while the decay approach infers the spin correlation statistically. 

While the two approaches that we have presented have different theoretical inputs and experimental observables, they reconstruct exactly the same quantum density matrix and consequently, all quantum quantities are measured to be the same. 
The kinematic approach involves the fewest possible kinematic observables to construct the density matrix event-by-event, and thus determines the entanglement event-by-event. The decay approach, on the other hand, measures more angular variables to infer the spin information statistically. The two approaches are thus complementary and offer important cross-checks of the same quantum information, in particular if there exists new physics that influences the production and decay differently. 

%%%%%%%%%%%%%%%%%%%%%%%%%%%
\customsection{Observability at the LHC}

%%%%%%%%%%%%%%%%%%%%%%%%
\subsection{Quantum Tomography in Drell-Yan Production $pp\to Z\to \mu^+\mu^-$}

With our kinematic approach formulation, the quantum information for rather simple processes can be studied, even for outgoing qubit particles that are stable.  One of the simplest and best-measured processes in hadronic collisions is the Drell-Yan production of a lepton pair.  In particular, the $Z$ boson provides one of the best laboratories with a $J=1$ ensemble.  

The spin correlation matrix of $\mu^+\mu^-$ produced from a $Z$ boson decay is given by Eq.~\eqref{eq:CijZmumu}. We obtain the concurrence and the Bell variable as
\begin{align}
\mathcal{C}(\Theta,\beta \approx 1) &= \frac{s_\Theta^2 (g_A^2-g_V^2)}{(1+c_\Theta^2)(g_A^2+g_V^2)}, \\
\mathcal{B}(\Theta,\beta \approx 1) &= \frac{2 \sqrt{2}\left(g_A^2+g_V^2 c_\Theta ^2\right)}{\left(1+c_\Theta ^2\right)\left(g_A^2+g_V^2\right)}, 
\end{align}
where $g_V$ and $g_A$ are, respectively, the vector and axial coupling strengths of the $Z$ boson to the charged leptons,\footnote{See Appendix~\ref{app:onshellZ} for the calculation of the variables $\mathcal{B}$, $\mathcal{C}$ and their variances.} and $\Theta$ is the scattering angle between the outgoing $\mu^-$ and the incoming beam in the $Z$ boson rest frame. 

\begin{figure}
  \centering
  \includegraphics[width=0.9\linewidth]{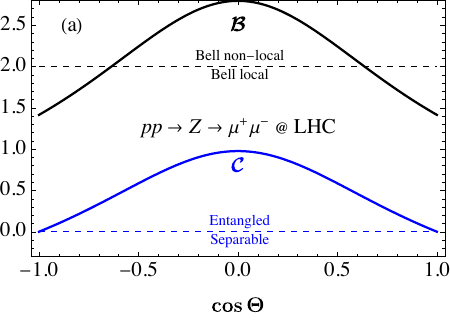}
  \vskip 2em
  \includegraphics[width=0.9\linewidth]{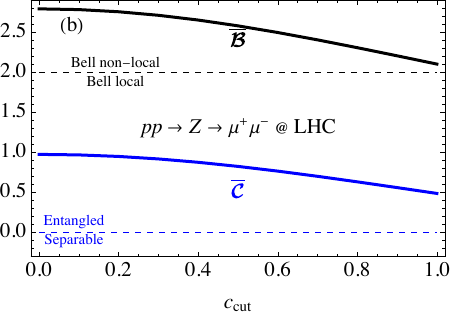}
  \caption{(a) Quantum tomography distribution for $Z\to \mu^+\mu^-$ at the LHC, (b) concurrence $\overline{\mathcal{C}}$ and the Bell variable $\overline{\mathcal{B}}$ integrated over angles satisfying $|\cos\Theta|<c_{\rm cut}$. The horizontal dashed lines indicate the boundaries of the entanglement and Bell non-locality.}
  \label{fig:LHCcut}
\end{figure}

We show the full quantum tomography for $Z \to \mu^+\mu^-$ at the LHC and calculate the concurrence and Bell variable in Fig.~\ref{fig:LHCcut}. In (a) for the angular distributions of $\mathcal{C}$ and $\mathcal{B}$, we see that the events with the largest quantum effects are in the central region. This is because the $\mu^+\mu^-$ system is mostly a spin-triplet with larger entanglement for $J_z=0$.  In (b), we calculate the integrated state with a selection cut of $|\cos\Theta|<c_{\rm cut}$ to better isolate the signals of entanglement and Bell non-locality.  The horizontal dashed lines denote the boundaries of Bell inequality violation and entanglement.

The experimental observability is determined by the reachable uncertainties $\Delta \overline{\mathcal{C}}^{ \rm stat}$ and $\Delta \overline{\mathcal{B}}^{{ \rm stat}}$, which depends on the statistics as given in Eq.~\eqref{eq:dvar}.  The variance, over the full phase space, is
\begin{equation}
    {\rm Var}(\mathcal{B})=2{\rm Var}(\mathcal{C}) = \frac{3(\pi-3)(g_A^2-g_V^2)^2}{2(g_A^2+g_V^2)^2}.
\end{equation}
The LHC has a large number of  $Z\to \mu^+\mu^+$ events~\cite{cmscollaboration2024measurementinclusivecrosssections} which will accumulate $10^8$ events from a luminosity of $300~ {\rm fb}^{-1}$ at Run 3.  The percentage accuracies, including only statistical uncertainty, of $\Delta \overline{\mathcal{C}}^{\rm stat}/\overline{\mathcal{C}}$ and $\Delta \overline{\mathcal{B}}^{\rm stat}/\overline{\mathcal{B}}$ can easily achieve $0.01\%$. 

%%%%%%%%%%%%%%%%%%%%%%%
\subsection{Quantum Tomography in $pp\to t\bar t$}

The ATLAS and CMS collaborations have observed entanglement in $t\bar t$ events~\cite{ATLAS:2023fsd,CMS:2024pts}; both using the decay approach in their analysis.  Here we present a comparative study between the kinematic approach and the decay approach for the $t\bar t$ system at the LHC.  Theoretically, both approaches will result in the same density matrix and thus the same quantum information. 

The $t\bar t$ system produced at the LHC is a mixed state including contributions from both $gg$ and $q\bar q$ initial states with a density matrix of
\begin{equation}
\rho(\Theta,\beta)=\frac{L_{q\bar q} |\mathcal{M}_{q\bar q\to t\bar t}|^2 \rho_{q\bar q \to t\bar t} +L_{gg} |\mathcal{M}_{gg\to t\bar t}|^2 \rho_{gg\to t\bar t}}{L_{q\bar q}|\mathcal{M}_{q\bar q\to t\bar t}|^2 +L_{gg}|\mathcal{M}_{gg\to t\bar t}|^2},
\end{equation}
where $L_{q\bar q}$ and $L_{gg}$ are the parton luminosities, respectively, of $q\bar{q}$ and $gg$. Unlike in the process $pp \to Z\to \mu^+\mu^-$, the variables $\mathcal{C}(\Theta,\beta)$ and $\mathcal{B}(\Theta,\beta)$ depend on the parton luminosities and require a numerical calculation.

\begin{figure}
  \centering
  \includegraphics[width=0.88\linewidth]{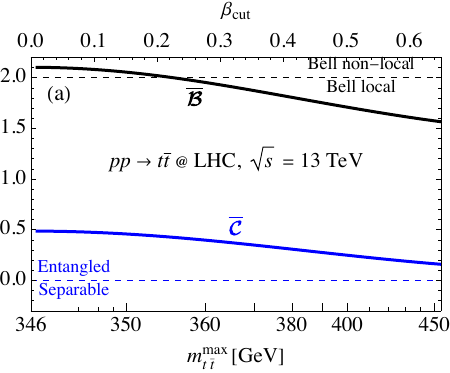}
  \includegraphics[width=0.88\linewidth]{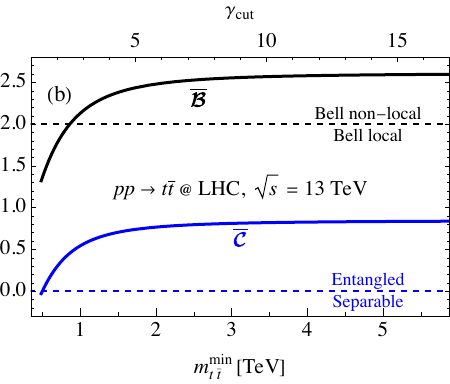}
  \caption{The concurrence $\mathcal{C}$ and the Bell variable $\mathcal{B}$ as a function of (a) upper invariant mass cut  $m_{t\bar t} < m_{t\bar t}^{\rm max}$; (b) lower invariant mass cut $m_{t\bar t}>m_{t\bar t}^{\rm min}$.  An angular cut $\cos\Theta<0.5$ is also imposed in (b) to select the high-$p_T$ region. The horizontal dashed lines indicate the boundaries of the entanglement and Bell non-locality violation.  The statistical uncertainty is listed in Table~\ref{tab:ppttUncertaity}.}
  \label{fig:ppttbetacut}
\end{figure}

The $t\bar t$ system is entangled near threshold, in a spin-singlet configuration, and also in the high-$p_T$ region, in a spin-triplet configuration.  The concurrence and the Bell variable at the LHC are shown in Fig.~\ref{fig:ppttbetacut}(a) near threshold with an upper $m_{t\bar t}$ cut, and in Fig.~\ref{fig:ppttbetacut}(b) in the boosted region with a lower cut on $m_{t\bar t}$.  Under the Standard Model prediction, these results are the same using the kinematic approach or the decay approach.

Where the difference between approaches arises is in the statistical uncertainties on the $\overline{\mathcal{C}}$ and $\overline{\mathcal{B}}$ measurements.  The statistical uncertainties are calculated in Table~\ref{tab:ppttUncertaity} at the LHC with an integrated luminosity of 300 fb$^{-1}$.  For both approaches, we include the fully leptonic $t\bar{t}$ branching ratio and the only selection cuts we apply are $m_{t\bar{t}} <350~{\rm GeV}$ in the threshold region and $m_{t\bar{t}} >1.5~{\rm TeV}$ and $|\cos\Theta|<0.5$.  The resulting statistical uncertainties using the decay approach are $1\% -26\%$.  In all cases the kinematic approach achieves a statistical uncertainty that is more than an order of magnitude smaller.\footnote{We do not include reconstruction efficiencies.  While the absolute uncertainties may change in a fully realistic simulation, the ratio between approaches is robust.}

\begin{table}[h]
    \centering
    \begin{tabular}{|c|c||c|c|c|c|}
    \hline
        & cuts & $\overline{\mathcal{C}}$ & $\Delta \overline{\mathcal{C}}^{\rm stat}$ & $\overline{\mathcal{B}}$ & $\Delta \overline{\mathcal{B}}^{\rm stat}$  \\
    \hline
        Kinematic & \multirow{2}{*}{$m_{t\bar t}<350\,$GeV} & \multirow{2}{*}{0.45}  & $1.0\times 10^{-4} $ & \multirow{2}{*}{2.04} & $1.4\times 10^{-4} $ \\
    \cline{1-1}\cline{4-4}\cline{6-6}
        Decay &   &  & $ 1.4 \times 10^{-2} $ &  & $ 6.3 \times 10^{-2} $ \\
    \hline
        Kinematic & $m_{t\bar t}>1.5\,$TeV & \multirow{2}{*}{0.70}  & $ 2.6\times 10^{-3}  $ & \multirow{2}{*}{2.37} &$ 3.8\times 10^{-3}  $ \\
    \cline{1-1}\cline{4-4}\cline{6-6}
        Decay & $|\cos\Theta|<0.5$   &  & $ 5.6 \times 10^{-2} $ & & 0.26 \\
    \hline
    \end{tabular}
    \caption{Statistical uncertainties on $\overline{\mathcal{C}}$ and $\overline{\mathcal{B}}$ measurements for $pp\to t\bar t$ with representative selection cuts. An integrated luminosity of 300 fb$^{-1}$ is used. The fully leptonic branching ratio of $t\bar{t}$ is included in both approaches without additional kinematic cuts.}
    \label{tab:ppttUncertaity}
\end{table}

%%%%%%%%%%%%%%%%%%%%%%%
\customsection{Discussion and Conclusions}

In studying the properties of the quantum states produced at colliders,  most efforts have focused on utilizing the decays of the particles to infer their spins.  We have shown that the production of the particles in $2\to 2$ processes provides sufficient information on the outgoing spins to perform quantum tomography without using the decays. 

One important advantage of the kinematic approach is that it enables using particles that do not decay in collider quantum tomography.  This opens up an entirely new class of processes at colliders for which we can study quantum information.  As a first step, we demonstrated this approach with the $pp \to Z\to \mu^+\mu^-$, which could be the first channel in which Bell non-locality is observed at the LHC.

Although we illustrated our results with the two-qubit system, the kinematic approach is also applicable to other quantum states such as two qutrit system $W^\pm$, as the spin density matrix is also determined by the two kinematic variable $\Theta$ and $\beta$~\cite{Hagiwara:1986vm}. This approach can be generalized to an $n$-body quantum system as well, in terms of the minimal number of kinematic variables.

 The kinematic approach also involves the fewer kinematic observables for particle pairs such as $t\bar t$ that decay.  This is because the kinematic approach only needs the total momentum of top decay products while the decay approach requires additional angular separations between top decay products. 
The two approaches are thus complementary and offer meaningful cross-checks of the same quantum information.  This can be useful, for example, if there exist beyond-the-Standard-Model predictions that modify the production and decay of the particles differently, then the decay approach and kinematic approach are sensitive to different modifications. 

Due to the simpler measurement and less demanding event reconstruction, the kinematic approach offers tremendous upside in reconstructing the quantum density matrix at colliders.  With the density matrix, many quantum quantities, for example quantum discord~\cite{Afik:2022dgh,Han:2024ugl}, EPR steering~\cite{Afik:2022dgh}, or magic~\cite{White:2024nuc}, can be calculated.  Furthermore, this method is sensitive to the effects from new physics beyond the Standard Model that may couple to the quantum state at production, such as a new pseudoscalar -- a CP-odd Higgs $A$ or a $t\bar t$ bound state $\eta_t$ -- or a vector -- a $Z'$ vector boson or a $t\bar t$ bound state $\Upsilon_t$ --  to modify the spin-singlet or spin-triplet predictions.

Our presentation has focused on statistical uncertainties.  In realistic projections, systematic uncertainty should be included.  Unless an experiment has sufficient data that their results are systematics dominated, reducing the statistical uncertainty is of critical importance. In a study of low energy $e^+ e^- \to \tau^+ \tau^-$ scattering, it was found that the reduced statistical uncertainty plays a crucial role in the observability even considering realistic systematic uncertainties~\cite{Han:2025ewp}.

%We note that our approach in Eq.~\eqref{eq:conc_kin} applies for quantities that are linear with respect to the spin correlation matrix elements.  For non-linear cases, one should measure the mean and variance of the spin correlation matrix elements themselves, to ensure consistency with the decay method.
Throughout our presentation, we illustrate our method for quantities that are linear with respect to the spin correlation matrix elements $C_{ij}$. For non-linear cases, one should first measure the mean and variance of the spin correlation matrix elements in accordance with the fictitious states, and then construct the non-linear variables, similar to the decay method.

In conclusion, we have proposed a new approach to explore the quantum information at colliders, the kinematic approach, which is based on $2 \to 2$ production kinematics. There is significant complementarity between this approach and the decay approach which uses particle decays to compute spin correlations.  Due to the simplicity of $2 \to 2$ kinematics and particle reconstruction, 
%\Tao{the kinematic approach has a lower statistical variance,  reaches a higher experimental acceptance and larger data sample, thus promises to}
the kinematic approach has a smaller statistical uncertainty.  It also benefits from easily including multiple decay channels leading to an effectively larger data sample, and thus
achieves the optimal statistical outcome for quantum tomography of the relevant processes and has a wide range of applicability.  In particular, we can move beyond the scope of using only promptly-decaying particles to using nearly the full array of observable particles in the Standard Model to construct the quantum information.

%%%%%%%%%%%%%%%%%%%%%%%%%%%%%%%%%%%%
\begin{acknowledgments}
%\vskip 1em
TH would like to thank the Aspen Center for Physics which is supported by the National Science Foundation (NSF) grant PHY-1607611, and the CERN TH Department for hospitality during the final stage of the project.  This work was supported in part by the U.S.~Department of Energy under grant No.~DE-SC0007914 and in part by the Pitt PACC. ML is also supported by the National Science Foundation under grant No.~PHY-2112829.
\end{acknowledgments}

\newpage
\appendix
\begin{widetext}
\section{Steps of the Kinematic Approach}
\label{app:onshellZ}

Using the kinematic approach requires process-specific input both in the measurement and in the prediction.  In this appendix we present the explicit steps to measure the concurrence and the Bell variable in the process $pp\to Z \to f\bar f$. 

The spin of the $Z$ boson along the beam direction can only be $\pm 1$ due to its vector couplings with massless initial states.  The amplitude of the $Z \to f\bar f$ decay is
\begin{equation} \label{eq:amplitude}
\mathcal{M}_{f\bar f}^{\pm} = \left(\frac{g}{\cos\theta_W}\right) \epsilon_\mu^{S_z=\pm 1} \Big(  g_V \bar f \gamma^\mu f + g_A \bar f \gamma_5\gamma^\mu f \Big),
\quad\quad\quad
\epsilon_\mu^{S_z=\pm 1}= (0,-1,\mp i,0),
\end{equation}
where $\theta_W$ is the Weinberg angle, $g_A$ and $g_V$ are the vector and axial couplings of $f\bar f$ currents to the $Z$ boson, with $g_V=-1/4 + \sin^2\theta_W$ and $g_A=-1/4$ for $e^+ e^-$, $\mu^+\mu^-$, or $\tau^+\tau^-$.

The squared matrix element summing over the spin of the final state is
\begin{equation}
 \sum|\calM^\pm|^2= 2 \left(\frac{g}{\cos\theta_W}\right)^2 \left(\beta^2 g_A^2 (1+c_\Theta^2) \pm 4\beta g_A g_V c_\Theta + g_V^2(2-\beta^2 s_\Theta^2) \right).
\end{equation}
The spin density matrix from the helicity amplitude is
\begin{equation}
    \rho_{\alpha\beta,\bar \alpha\bar\beta} = \frac{\mathcal{M}_{f_\alpha f_{\bar \alpha}}\mathcal{M}^*_{f_{\beta}f_{\bar \beta}}}{\sum|\mathcal{M}|^2} ,
\end{equation}
and the spin correlation matrix $C_{ij} = \tr(\rho \sigma_i\otimes \sigma_j)$ is 
\begin{align}\label{eq:Cijmumu}
&C_{ij}^{S_z=\pm 1} = \nn
&\left(\begin{array}{ccc}
\frac{g_V^2(2c_\Theta^2+\beta^2 s_\Theta^2) + g_A^2\beta^2(1+c_\Theta^2)\pm 4g_V g_A\beta}{\beta^2 g_A^2 (1+c_\Theta^2) \pm 4\beta g_A g_V c_\Theta + g_V^2(2-\beta^2s_\Theta^2)}
& 0 & 
- \frac{2 g_V s_\Theta \sqrt{1-\beta^2}(g_V c_\Theta \pm g_A \beta)}{\beta^2 g_A^2 (1+c_\Theta^2) \pm 4\beta g_A g_V c_\Theta + g_V^2(2-\beta^2 s_\Theta^2)}\\
0 & \frac{(g_A^2-g_V^2)\beta^2 s_\Theta^2}{\beta^2 g_A^2 (1+c_\Theta^2) \pm 4\beta g_A g_V c_\Theta + g_V^2(2-\beta^2s_\Theta^2)} & 0 \\
-\frac{2 g_V s_\Theta \sqrt{1-\beta^2}(g_V c_\Theta \pm g_A \beta)}{\beta^2 g_A^2 (1+c_\Theta^2) \pm 4\beta g_A g_V c_\Theta + g_V^2(2-\beta^2s_\Theta^2)} & 0 & \frac{s_\Theta^2((2-\beta^2)g_V^2 - \beta^2 g_A^2 )}{\beta^2 g_A^2 (1+c_\Theta^2) \pm 4\beta g_A g_V c_\Theta + g_V^2(2-\beta^2 s_\Theta^2)} 
\end{array}\right).
\end{align}
The spin correlation matrices $C_{ij}^{S_z=1}$ and $C_{ij}^{S_z=-1}$ are related by $g_A \leftrightarrow -g_A$. 

In $pp$ collisions the symmetric initial state means that the parton luminosities do not contribute to the weighting of different polarizations.  The spin correlation matrix summing over $Z$ boson polarizations is
\begin{equation}
    C_{ij}=\frac{|\mathcal{M}^+|^2 C_{ij}^{S_z=1} + |\mathcal{M}^-|^2 C_{ij}^{S_z=-1}}{|\mathcal{M}^+|^2+|\mathcal{M}^-|^2}.
\end{equation}
When $\beta=1$, both the cross section and the correlation matrix reduce to a simple form:
\begin{equation}\label{eq:CijZmumu}
    \frac{1}{\sigma}\frac{d\sigma}{d\cos\Theta}=\frac{3}{8}(1+\cos^2\Theta),~~
    C_{ij} = \begin{pmatrix}
        1 & 0 & 0 \\
        0 &  \frac{s_\Theta^2(g_A^2-g_V^2)}{(1+c_\Theta^2)(g_A^2+g_V^2)} & 0 \\
        0&0& -\frac{s_\Theta^2(g_A^2-g_V^2)}{(1+c_\Theta^2)(g_A^2+g_V^2)}
    \end{pmatrix}.
\end{equation}
The polarization vectors $B^\pm_i$ are a constant along the momentum $\vec k$ direction
\begin{equation}
    B_{i}^\pm = \left(\frac{2g_A g_V}{g_A^2+g_V^2},~ 0,~0 \right).
\end{equation}
The full quantum tomography for $Z \to \mu^+\mu^-$ at the LHC is shown in Fig.~\ref{fig:CijZmumu}.

\begin{figure}
  \centering
  \includegraphics[width=0.4\linewidth]{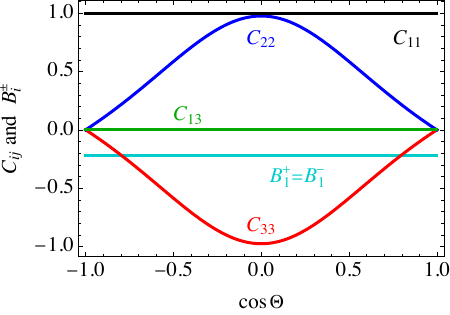}
  \caption{Complete quantum tomography distribution for $Z\to \mu^+\mu^-$ at the LHC.  All the other parameters that are not shown are zero. }
  \label{fig:CijZmumu}
\end{figure}

The concurrence and the Bell variable are  calculated to be 
\begin{align}
\mathcal{C}(\Theta,\beta=1) &= \frac{1}{2}(C_{22}-C_{33} + C_{11}-1) = \frac{s_\Theta^2 (g_A^2-g_V^2)}{(1+c_\Theta^2)(g_A^2+g_V^2)}, \\
    \mathcal{B}(\Theta,\beta=1) &= \sqrt{2}(C_{11}+C_{22}) = \frac{2 \sqrt{2}\left(g_A^2+g_V^2 c_\Theta ^2\right)}{\left(1+c_\Theta ^2\right)\left(g_A^2+g_V^2\right)}.
\end{align}
and their variances $\mathrm{Var}(\mathcal{R})
=\left\langle \left(\mathcal{R}-\overline{\mathcal{R}}\right)^2 \right\rangle
=\left\langle \mathcal{R}^2\right\rangle-\overline{\mathcal{R}}^2$ are calculated as follows,
\begin{subequations}
\begin{align}
    \overline{ \mathcal{C} } &= \frac{1}{\sigma} \int_{-1}^1 \frac{d\sigma}{d\cos\Theta} \mathcal{C}(\Theta,\beta=1)  ~d\cos\theta = \frac{g_V^2-g_A^2}{2(g_V^2+g_A^2)} \\
    \braket{ \mathcal{C}^2 } &= \frac{1}{\sigma} \int_{-1}^1 \frac{d\sigma}{d\cos\Theta} \left(\mathcal{C}(\Theta,\beta=1)\right)^2  ~d\cos\theta = \frac{3\pi - 8}{4} \frac{(g_V^2-g_A^2)^2}{(g_V^2+g_A^2)^2} \\
    \Longrightarrow~ \mathrm{Var}(\mathcal{C}) &= \frac{3(\pi-3)}{4}\frac{(g_V^2-g_A^2)^2}{(g_V^2+g_A^2)^2}
\end{align}
\end{subequations}
and
\begin{subequations}
\begin{align}
    \overline{ \mathcal{B} } &= \frac{1}{\sigma} \int_{-1}^1 \frac{d\sigma}{d\cos\Theta} \mathcal{B}(\Theta,\beta=1)  ~d\cos\theta = \frac{g_V^2+3g_A^2}{\sqrt{2}(g_V^2+g_A^2)} \\
    \braket{ \mathcal{B}^2 } &= \frac{1}{\sigma} \int_{-1}^1 \frac{d\sigma}{d\cos\Theta} \left(\mathcal{B}(\Theta,\beta=1)\right)^2  ~d\cos\theta =  \frac{3\pi g_A^4+6(4-\pi)g_V^2 g_A^2+ (3\pi -8 ) g_V^4}{2(g_V^2+g_A^2)^2} \\
    \Longrightarrow~ \mathrm{Var}(\mathcal{B}) &= \frac{3(\pi-3)}{2}\frac{(g_V^2-g_A^2)^2}{(g_V^2+g_A^2)^2} = 2\mathrm{Var}(\mathcal{C}).
\end{align}
\end{subequations}

\end{widetext}

\bibliographystyle{apsrev4-1}
\bibliography{refs}
\end{document}